\documentclass[11pt,a4paper]{article}
\usepackage{float}
\usepackage{multirow}
\usepackage{graphics}
\usepackage[cp1252,utf8]{inputenc}
\usepackage{hyperref}
\usepackage{graphicx} 
\usepackage{amsmath}
\usepackage{amsfonts}
\usepackage{amssymb}
\usepackage{bigints} 
\usepackage{relsize} 
\usepackage[top=1in, bottom=2cm, left=1.5cm, right=1.5cm]{geometry}
\usepackage{authblk}

\usepackage{changepage} 

\usepackage{afterpage}

\usepackage{placeins}

\usepackage{wrapfig}

\title{\bf Holographic entanglement entropy and generalized entanglement temperature}
\author[a]{\bf  Ashis Saha \thanks{sahaashis0007@gmail.com, ashisphys18@klyuniv.ac.in}}
\author[b]{\bf Sunandan Gangopadhyay \thanks{sunandan.gangopadhyay@gmail.com, sunandan.gangopadhyay@bose.res.in}}
\author[a]{\bf Jyoti Prasad Saha \thanks{jyotiprasadsaha@gmail.com}}

\affil[a]{\textit{Department of Physics, University of Kalyani, Kalyani 741235, India}}

\affil[b]{\textit{Department of Theoretical Sciences, S.N.~Bose National Centre for Basic Sciences,}
\textit{JD Block, Sector-III, Salt Lake, Kolkata 700106, India}}

\date{}

\begin{document}

\maketitle

\begin{abstract}
\noindent In this work we study the flow of holographic entanglement entropy in dimensions $d \geq 3$ in the gauge/gravity duality set up. We observe that a generalized entanglement temperature $T_g$ can be defined which gives the Hawking temperature $T_H$ in the infrared region and leads to a generalized thermodynamics like law $E= \left(\frac{d-1}{d}\right)T_g~S_{REE}$, which becomes an exact relation in the entire region of the subsystem size $l$, including both the infrared ($l\rightarrow\infty$) as well as the ultraviolet ($l\rightarrow 0$) regions. Furthermore, in the IR limit, $T_g$ produces the Hawking temperature $T_H$ along with some correction terms which bears the signature of short distance correlations along the entangling surface. Moreover, for $d\geq 3$, the IR limit of the renormalized holographic entanglement entropy gives the thermal entropy of the black hole as the leading term, however, does not have a logarithmic correction to the leading term unlike the BTZ black hole ($d=2$) case. The generalized entanglement temperature $T_g$ also firmly captures the quantum mechanical to thermal crossover in the dual field theory at a critical value $l_c$ of the subsystem size in the boundary which we graphically represent for $AdS_{3+1}$ and $AdS_{4+1}$ black holes. We observe that this critical value $l_c$ where the crossover takes place decreases with increase in the dimension of the spacetime.   
\end{abstract}

\section*{Introduction}
The von Neumann entropy or the entanglement entropy (EE) is one of the fundamental and wellstudied entities of quantum physics \cite{nielsen}. It is non-local in nature and represents how the degrees of freedom of two subsystems are correlated. This bipartite scenario can be extended further to a multipartite scenario \cite{multi}. In the bipartite case, there exists two subsystems namely $A$ and $B$. The mathematical definition of the EE is then given by
\begin{eqnarray}
S_{A} = -tr[\rho_A \log\rho_A]
\end{eqnarray}
where $\rho_A$ is the reduced density matrix of the subsystem $A$ constructed by tracing out the degrees of freedom of the subsystem $B$ from the total density matrix ($\rho_{total}$) of the system as
\begin{eqnarray}
\rho_A = tr_B [\rho_{total}].
\end{eqnarray}
The EE has been a matter of great interest in quantum information theory as it captures the amount of information loss suffered due to spatial division of the concerned system. It has also been shown a strong connection exists relating the quantum information with the thermal entropy of a system \cite{eetht}. This led to the computation of EE in the field theoretical set up. However, The computation of EE in field theoretical scenario is notoriously difficult as it is divergent in nature and demands a number of symmetries for analytical computations. In a field theory with conformal group symmetry, the EE has been computed in $1+1$-dimensions for various topologies of the subsystems \cite{Calabrese}. However, the computation of EE in a interacting field theory or a higher dimensional free field theory is not quite clear \cite{freefield}. This problem was resolved by the $AdS/CFT$ correspondence in a satisfactory way. \cite{maldacena}, \cite{maldacena2}. The interesting connection between a weakly coupled gravitational theory and a strongly coupled quantum field theory (QFT), provided by the $AdS$/CFT correspondence has been a matter of great interest for the last two decades \cite{witten}, \cite{susskind}. This correspondence has led to the concept of holographic computation of EE via the Ryu-Takayanagi prescription \cite{RT_prl}. The holographic computation of EE states that the holographic entanglement entropy (HEE) of a asymptotically $AdS_{d+1}$ spacetime is equal to EE of a $d$-dimensional CFT which lives at the boundary of the classical bulk theory. The formula for the HEE is given by \cite{RT_jhep}
\begin{eqnarray}
S_{HEE} = \frac{Area(\gamma_A)}{4G_{d+1}}
\end{eqnarray}
where $\gamma_A$ represents the static minimal surface extending into the bulk due to the presence of the subsystem $A$ at the boundary of the assymptotically $AdS$ spacetime. Interestingly, this is quite similar to the famous Bekenstein-Hawking area law for black hole entropy \cite{bhe}. The computation of HEE under different circumstances has been a well studied subject thereafter \cite{sk1}-\cite{cpark2}. The computation of HEE has also been extended to time-dependent scenarios. This is the covariant HEE prescription \cite{HRT}. Interestingly in the UV limit, the HEE gives a thermodynamics like law. This has been named as entanglement thermodynamics in the literature which in turn gives a entanglement temperature $T_{ent}$ \cite{jprl}. However, to define the entanglement temperature one has to neglect the higher order terms in the temperature expansion of the HEE \cite{alishahiha}. This restricts the definition of $T_{ent}$ only in the UV domain ($\frac{l}{z_h}\ll 1$) of the theory. One must note that the definition of $T_{ent}$ is also not quite clear as it is quite different from our known notion of a thermodynamic temperature and it is inversely proportional to the the subsystem size at the boundary. This invokes the thrust to look out for a real thermodynamical notion of a temperature in the theory which shall have a scale independent, global defintion unlike $T_{ent}$, as this would  provide a deep understanding about the microscopic origin of thermodynamics. Moreover, this raises another question about  a possible connection existing between the entanglement temperature and a real thermodynamical temperature.\\
The usual approach to the computation of the HEE has been heavily influenced by the UV and IR domain of the theory determined by the size of the boundary subsystem. However, to observe the evolution of the HEE with the subsystem size correctly one has to work with the entire domain of subsystem size $l$. In case of a ($1+1$)-dimensional conformal field theory (CFT) this is quite simple as the computed results of EE has a well defined closed analytical form. This leads to some fascinating results by providing us a connection between EE and Hawking entropy along with the Hawking temperature \cite{cpark3}, but the story drastically changes in higher dimensions due to technical computations.\\ 
In this work we observe the evolution of HEE with the subsystem size in dimensions $d \geq 3$ in both UV and IR domain of the theory by incorporating all the terms in the expansion. Such a study was carried out earlier in \cite{cpark3} in the case of the $2+1$-dimensional BTZ black hole which corresponds to a $1+1$-dimensional CFT. We define a generalized entanglement temperature $T_g$ using the first law of black hole thermodynamics ($E= \left(\frac{d-1}{d}\right)T_H S_{BH}$) and we show that $T_g$ produces the exact Hawking temperature in the IR domain of the dual field theory. We specifically represent this for $AdS_{3+1}$ and $AdS_{4+1}$ black holes. These interesting results indicate that the entanglement temperature can evolve to a real thermodynamic temperature satisfying a real thermodynamical law. In the ultraviolet domain, $T_g$ gives rise to an entanglement temperature $T_{e}$ that once again satisfies the thermodynamics like law $E = \left(\frac{d-1}{d}\right)T_{e} S_{REE}^{(UV)}$. It should be noted that the generalized entanglement temperature defined in this paper should not be identified as the thermodynamic temperature of the boundary CFT. The thermodynamic temperature of the CFT (which does not flow with the subsystem size $l$) corresponds to the Hawking temperature of the black hole and arises from the IR limit of generalized entanglement temperature $T_g$. We would like to emphasize that the definition of the generalized entanglement temperature given in this paper is different from the definition of the entanglement temperature given in \cite{jprl} upto a constant factor. It should be noted that the definition given in \cite{jprl} agrees with the Hawking temperature only upto a constant multiplicative factor. We have demonstrated this explicitly for the AdS$_{3+1}$ and BTZ black holes.\\
The paper is organized as follows. In section \ref{sec1}, we holographically compute the EE of both the ground state and the excited state of a $d$-dimensional field theory by incorporating the Ryu-Takayanagi conjecture for a strip like subsystem. We then produce a divergence free entanglement entropy, namely, renormalized entanglement entropy $S_{REE}$. In section \ref{gent}, we define a generalized entanglement temperature $T_g$ by utilizing the first law of black hole thermodynamics and the result of $S_{REE}$. We also study the behavior of $T_g$ in the UV and IR domain in arbitrary dimensions. In section \ref{sec3}, we concentrate on $AdS_{3+1}$ and $AdS_{4+1}$ black holes to graphically represent the variation of the inverse generalized entanglement temperature $\beta_g$ with respect to the subsystem size $l$. We then observe the variation of the dimensionless quantity $\frac{d\left(\frac{\beta_g}{z_h}\right)}{d\left(\log \left(\frac{l}{z_h}\right)\right)}$ against $\frac{l}{z_h}$. Finally, we discuss the results that we have obtained in section \ref{sec4}. The paper ends with an appendix.

\section{Renormalized holographic entanglement entropy}\label{sec1}
In this section, We shall compute the renormalized HEE for the AdS Schwarzschild black hole in $d+1$-dimensions. To begin with we write down the metric of the $d+1$-dimension AdS Schwarzschild black hole

\begin{eqnarray}\label{d+1}
ds^2 = \frac{R^2}{z^2} \bigg[-f(z)^2dt^2+\frac{dz^2}{f(z)^2}+\sum_{k=1}^{d-1} dx_{k}^2\bigg]~.
\end{eqnarray}
This geometry is then dual to a $d$-dimensional CFT at the boundary by the AdS/CFT correspondence. The expression for the lapse function $f(z)$ contains the information about the excitation properties of the boundary CFT.
The ground state in the $d$-dimensional boundary CFT corresponds the lapse function $f(z)=1$ and the metric represents pure $AdS$ spacetime in $(d+1)$-dimensions:
\begin{eqnarray}\label{pureads}
ds^2 = \frac{R^2}{z^2} \bigg[-dt^2+dz^2+\sum_{k=1}^{d-1} dx_{k}^2\bigg]~.
\end{eqnarray}
Similarly, a thermally excited state in the $d$-dimensional boundary CFT leads to a deformed $AdS$ spacetime. More precisely, the excited state in the CFT corresponds to an assymptotically $AdS$ space with a black hole with the lapse function 
\begin{eqnarray}
f(z)=\sqrt{1-\frac{z^d}{z_h^d}}~.
\end{eqnarray}
where $z_h$ is the event horizon radius of the black hole. The metric is the well known $AdS_{d+1}$ Schwarzschild black hole geometry and is given by
\begin{eqnarray}\label{sads}
ds^2 = \frac{R^2}{z^2} \bigg[(1-\frac{z^d}{z_h^d})dt^2+\frac{dz^2}{(1-\frac{z^d}{z_h^d})}+\sum_{k=1}^{d-1} dx_{k}^2\bigg]~.
\end{eqnarray}
The Hawking temperature of the black hole reads
\begin{eqnarray}\label{hawkingt}
T_H=\frac{d}{4\pi z_h}~.
\end{eqnarray}
\noindent We now proceed to holographically calculate the entanglement entropy of the excited state of the $d$-dimensional boundary CFT. 
\subsection{Holographic entanglement entropy of the $AdS_{d+1}$ Schwarzschild black hole}
To begin our analysis, we first make the choice of subsystem $A$ at the boundary. This is a strip whose geometry is specified as $-\frac{l}{2} < x_1 < +\frac{l}{2}$ and $-\frac{L}{2} < x_{2,3,4,..,d-1} < +\frac{L}{2}$. This specifies the volume of the subsystem at the boundary field theory to be
\begin{eqnarray}
V_{(A)}= L^{d-2}l~.
\end{eqnarray}
The thermal entropy of the boundary field theory which is the amount of Bekenstein-Hawking entropy contained in the above mentioned boundary subsystem volume is given by \cite{cpark3}
\begin{eqnarray}\label{hawking}
S_{BH} = \frac{1}{4G_{d+1}}\frac{L^{d-2}l}{z_h^{d-1}}
\end{eqnarray}
where, $G_{d+1}$ is the Newton's gravitational constant in $(d+1)$-dimensions.\\
\noindent On the First step, we proceed to compute the HEE and the corresponding subsystem size in terms of the bulk coordinates of the metric (\ref{sads}). Here we keep $L$, representing the width of the strip is fixed but the length of the strip denoted by $l$ can vary from zero to infinity.  We parametrize the static minimal surface $\gamma_A$ by $x_1=x(z)$.\\
\noindent The corresponding area of $\gamma_A$ reads
\begin{eqnarray}\label{EE}
	A(\gamma_A) &=& 2 R^{d-1} L^{d-2} \int_{a}^{z_t} \frac{z_t^{d-1} dz}{z^{d-1}\sqrt{1-\frac{z^d}{z_h^d}}\sqrt{z_t^{2(d-1)}-z^{2(d-1)}}} \nonumber\\
\nonumber	\\
	&=& \frac{2 R^{d-1} L^{d-2}}{z_t^{(d-2)}} \sum_{n=0}^{\infty}\frac{\Gamma[n+\frac{1}{2}]}{\sqrt{\pi}\Gamma[n+1]} \alpha^{nd}\int_{(a/z_t)}^{1} du \frac{u^{nd}}{u^{(d-1)}\sqrt{1-u^{2(d-1)}}};~~ u= \frac{z}{z_t},\alpha= \frac{z_t}{z_h}
\end{eqnarray}
The holographic entanglement entropy ($S_{E}$) can now be computed from the RT formula \cite{RT_prl}
\begin{eqnarray}\label{excited}
S_{E} &=&\frac{A(\gamma_A)}{4G_{d+1}}\nonumber\\
&=& \frac{1}{2(d-2)G_{d+1}} \bigg(\frac{L}{a}\bigg)^{(d-2)} - \frac{\sqrt{\pi}}{2G_{d+1}} \bigg(\frac{L}{z_t}\bigg)^{(d-2)} \frac{\Gamma[\frac{d}{2(d-1)}]}{(d-2)\Gamma[\frac{1}{2(d-1)}]}\nonumber\\
&& +\frac{1}{2G_{d+1}}\left(\frac{L}{z_t}\right)^{(d-2)} \sum_{n=1}^{\infty} \frac{\Gamma[n+\frac{1}{2}]}{\Gamma[n+1]} \frac{\Gamma[\frac{2+(n-1)d}{2(d-1)}]}{2(d-1)\Gamma[\frac{1+nd}{2(d-1)}]} \bigg(\frac{z_t}{z_h}\bigg)^{nd}; ~ (d \geq 3)~.
\end{eqnarray}
$z_t$ is the turning point in the bulk satisfying the condition $\frac{dz}{dx_1}|_{z=z_t}=0$. It is the maximal value of $z$ on minimal surface $\gamma_A$. The position of the turning point $z_t$ with respect to the event horizon radius $z_h$ holographically determines the UV ($\frac{z_t}{z_h}\ll1$) and IR ($\frac{z_t}{z_h}\approx 1$) domains of the concerned field theory from the gravity side of the story.\\
 \noindent The length ($l$) of the boundary subsystem $A$ in terms of the turning point $z_t$ reads

\begin{eqnarray}\label{system}
l &=& 2 \int_{0}^{z_t} \frac{z^{(d-1)} dz}{z_t^{(d-1)} \sqrt{1-\frac{z^d}{z_h^d}}\sqrt{1-(\frac{z}{z_t})^{2(d-1)}}}\nonumber\\
\nonumber\\
&=& 2z_t \sqrt{\pi} \frac{\Gamma[\frac{d}{2(d-1)}]}{\Gamma[\frac{1}{2(d-1)}]} +\sum_{n=1}^{\infty} \bigg(\frac{2z_t}{1+nd}\bigg)\frac{\Gamma[n+\frac{1}{2}]}{\Gamma[n+1]}\frac{\Gamma[\frac{d(1+n)}{2(d-1)}]}{\Gamma[\frac{1+nd}{2(d-1)}]} \bigg(\frac{z_t}{z_h}\bigg)^{nd}~.
\end{eqnarray}
 In the above expressions we have set the $AdS$ radius $R=1$.
\subsection{Holographic entanglement entropy of the pure $AdS_{d+1}$ spacetime}
\noindent We move onto compute the holographic renormalized entanglement entropy of the $AdS_{d+1}$ Schwarzschild black hole. This we shall do by substracting the holographic entanglement entropy of the pure $AdS_{d+1}$ spacetime from the holographic entanglement entropy of the $AdS_{d+1}$ Schwarzschild black hole. The holographic entanglement entropy $(S_{G})$ and subsystem size $l$ of the pure $AdS_{d+1}$ spacetime reads \cite{RT_jhep}
\begin{eqnarray}\label{ground}
S_{G} = \frac{1}{2(d-2)G_{d+1}} \bigg(\frac{L}{a}\bigg)^{(d-2)} - \frac{2^{(d-3)}\pi^{\frac{d-1}{2}}}{(d-2)G_{d+1}}\bigg(\frac{\Gamma[\frac{d}{2(d-1)}]}{\Gamma[\frac{1}{2(d-1)}]}\bigg)^{(d-1)} \bigg(\frac{L}{l}\bigg)^{(d-2)}
\end{eqnarray}
and
\begin{eqnarray}
l = 2\sqrt{\pi}\frac{\Gamma[\frac{d}{2(d-1)}]}{\Gamma[\frac{1}{2(d-1)}]}~z_t^{(g)}
\end{eqnarray}
where $z_t^{(g)}$ is the turning point corresponding to ground state of the boundary CFT.\\
\noindent The relation between the turning point of the $AdS_{d+1}$ Schwarzschild geometry and that in the pure $AdS_{d+1}$ geometry is as follows\\
\begin{eqnarray}
z_t^{(g)}=z_t \bigg[1+ \frac{1}{\sqrt{\pi}}\sum_{n=1}^{\infty} \bigg(\frac{1}{1+nd}\bigg)\bigg(\frac{\Gamma[\frac{1}{2(d-1)}]}{\Gamma[\frac{d}{2(d-1)}]}\bigg)\frac{\Gamma[n+\frac{1}{2}]}{\Gamma[n+1]}\frac{\Gamma[\frac{d(1+n)}{2(d-1)}]}{\Gamma[\frac{1+nd}{2(d-1)}]} \bigg(\frac{z_t}{z_h}\bigg)^{nd}~\bigg]~.
\end{eqnarray} 
\noindent We now define a divergence free holographic entanglement entropy. This has been named as the renormalized holographic entanglement entropy $S_{REE}$ in the literature \cite{cpark3}, in $2+1$-dimensions. This is obtained by substracting the ground state entanglement entropy from the entanglement entropy for the excited state thereby removing the $l$ independent UV divergence term which arises due to short distance correlation along the entangling surface of the dual field theory. The renormalized holographic entanglement entropy in $d+1$-dimensions therefore reads \footnote{One can also this object as renormalized generalized holographic entanglement entropy since this object also receives thermal corrections in the IR limit. } 
\begin{eqnarray}\label{renorm}
S_{REE} &=& S_{E} - S_{G} \nonumber\\
&=&\frac{2^{(d-3)}\pi^{\frac{d-1}{2}}}{(d-2)G_{d+1}}\bigg(\frac{\Gamma[\frac{d}{2(d-1)}]}{\Gamma[\frac{1}{2(d-1)}]}\bigg)^{(d-1)} \bigg(\frac{L}{l}\bigg)^{(d-2)} - \frac{\sqrt{\pi}}{2G_{d+1}} \bigg(\frac{L}{z_t}\bigg)^{(d-2)} \frac{\Gamma[\frac{d}{2(d-1)}]}{(d-2)\Gamma[\frac{1}{2(d-1)}]}\nonumber\\
&& +\frac{1}{2G_{d+1}}\left(\frac{L}{z_t}\right)^{(d-2)} \sum_{n=1}^{\infty} \frac{\Gamma[n+\frac{1}{2}]}{\Gamma[n+1]} \frac{\Gamma[\frac{2+(n-1)d}{2(d-1)}]}{2(d-1)\Gamma[\frac{1+nd}{2(d-1)}]} \bigg(\frac{z_t}{z_h}\bigg)^{nd}~.
\end{eqnarray}
These expressions are valid for any dimension $d\geq 3$.\\

\section{A generalized entanglement temperature $T_g$}\label{gent}
In this section we proceed to define a generalized entanglement temperature in order to understand the behaviour of the holographic renormalized entanglement entropy ($S_{REE}$) for $d$-dimensional CFT along the entire domain of $l$. It has been realized in the case of the BTZ black hole that applying the first law of black hole thermodynamics, the UV behaviour of $S_{REE}$ leads to an entanglement temperature that is quite different from the Hawking temperature of the black hole. Hence, a generalized entanglement temperature was defined in \cite{cpark3} which captured both the UV and the IR regions in the $d=2$ case corresponding to the BTZ black hole. We shall do this for the $AdS_{d+1}$ Schwarzschild black hole in this investigation.\\

\noindent The first law of black hole thermodynamics reads \cite{BCH}
\begin{eqnarray}
dE = T_{H} ~dS_{BH}
\end{eqnarray}
which leads to the following change in the internal energy due to the thermal deformation of the pure $AdS_{d+1}$ spacetime  
\begin{eqnarray}\label{cE}
E &=& \int_{\infty}^{z_h} T_{H} dS_{BH} \nonumber\\
&=& \frac{(d-1)L^{(d-2)}l}{16\pi G_{d+1} z_h^d}~.
\end{eqnarray}
This internal energy represents the amount of information stored in the boundary volume. This information behaves in a quantum mechanical way in the UV region $\frac{l}{z_h}\ll1$ and in a thermal way in the IR region $l\rightarrow\infty$. Note that eq.(s)(\ref{hawkingt},\ref{hawking},\ref{cE}) yields the relation
\begin{eqnarray}
E = \left(\frac{d-1}{d}\right) T_H~S_{BH}~.
\end{eqnarray} 
 The expression for the internal energy given in eq.(\ref{cE}) along with the above thermodynamic relation, motivates us to define a generalized entanglement temperature $T_{g}$ as
\begin{eqnarray}\label{betagen}
\beta_{g}&\equiv&\frac{1}{T_{g}}=\frac{(d-1)}{d}\frac{S_{REE}}{E}~.
\end{eqnarray}
\noindent Substituting the expressions for $S_{REE}$ from eq.(\ref{renorm}) and $E$ from eq.(\ref{cE}), we obtain the expression of the generalized entanglement temperature $T_g$ for the $AdS_{d+1}$ black hole to be
\begin{eqnarray}
\frac{1}{T_{g}}&=& \frac{16\pi z_h^d}{ld} \bigg[\frac{2^{(d-3)}\pi^{\frac{d-1}{2}}}{(d-2)\Gamma[\frac{1}{2(d-1)}]}\bigg(\frac{\Gamma[\frac{d}{2(d-1)}]}{\Gamma[\frac{1}{2(d-1)}]}\bigg)^{(d-1)} \bigg(\frac{1}{l}\bigg)^{(d-2)}\nonumber\\
&&- \frac{\sqrt{\pi}}{2} \bigg(\frac{1}{z_t}\bigg)^{(d-2)} \frac{\Gamma[\frac{d}{2(d-1)}]}{(d-2)\Gamma[\frac{1}{2(d-1)}]}
+\frac{1}{2} \left(\frac{1}{z_t}\right)^{(d-2)} \sum_{n=1}^{\infty} \frac{\Gamma[n+\frac{1}{2}]}{\Gamma[n+1]} \frac{\Gamma[\frac{2+(n-1)d}{2(d-1)}]}{2(d-1)\Gamma[\frac{1+nd}{2(d-1)}]} \bigg(\frac{z_t}{z_h}\bigg)^{nd}\bigg].
\end{eqnarray}
This generalized entanglement temperature $T_g$ interpolates between the entanglement temperature in the UV region and a real thermodynamical temperature in the IR region. In the following section we shall verify this statement in spatial dimensions $d\geq 3$.
Note that we have defined the generalized entanglement temperature by putting a factor of $\left(\frac{d-1}{d}\right)$ in front of $\frac{S_{REE}}{E}$. The factor has been chosen so that $T_g$ goes to the Hawking temperature of the black hole in the IR ($l\rightarrow\infty$) limit. We also stress that this definition of the generalized entanglement temperature is different from that given in \cite{jprl}, namely, 
\begin{eqnarray}\label{bhattacharyya}
T_{ent}\Delta S=\Delta E
\end{eqnarray}
where $\Delta E$ is the increased amount of energy in the subsystem, given by
\begin{eqnarray}
\Delta E =  \frac{(d-1)L^{(d-2)}ml}{16\pi G_{d+1}} ~;~m=\frac{1}{z_h^d}~.
\end{eqnarray}
The above definition for the entanglement temperature does not lead to the exact Hawking temperature in the IR limit but is equal to the Hawking temperature upto a multiplicative constant factor. This we shall now demonstrate for the SAdS$_4$ black hole case. For the SAdS$_4$ black hole spacetime ($d=3$), the renormalized entanglement entropy reads
\begin{eqnarray}\label{d3}
S_{REE}= \left(\frac{\pi L}{G l}\right)\left(\frac{\Gamma[\frac{3}{4}]}{\Gamma[\frac{1}{4}]}\right)^2- \frac{\sqrt{\pi}}{2G}\left(\frac{L}{z_t}\right)\left(\frac{\Gamma[\frac{3}{4}]}{\Gamma[\frac{1}{4}]}\right)+\frac{1}{8G}\left(\frac{L}{z_t}\right)\sum_{n=1}^{\infty} \frac{\Gamma[n+\frac{1}{2}]}{\Gamma[n+1]}\frac{\Gamma[\frac{2+3(n-1)}{4}]}{\Gamma[\frac{1+3n}{4}]}\left(\frac{z_t}{z_h}\right)^{3n}~.
\end{eqnarray}
Using the definition for the entanglement temperature given in eq.(\ref{bhattacharyya}), and $\Delta S\equiv S_{REE}$ (given in eq.(\ref{d3})) with the expression for $\Delta E$ in $d=3$ spatial dimensions
\begin{eqnarray}
\Delta E= \frac{Ll}{8\pi G_4 z_h^3}
\end{eqnarray} 
we obtain the entanglement temperature to be 
\begin{eqnarray}
\frac{1}{T_{ent}}= \frac{8\pi z_h^3}{l}\left[\left(\frac{\pi }{ l}\right)\left(\frac{\Gamma[\frac{3}{4}]}{\Gamma[\frac{1}{4}]}\right)^2- \frac{\sqrt{\pi}}{2z_t}\left(\frac{\Gamma[\frac{3}{4}]}{\Gamma[\frac{1}{4}]}\right)+\left(\frac{1}{8z_t}\right)\sum_{n=1}^{\infty} \frac{\Gamma[n+\frac{1}{2}]}{\Gamma[n+1]}\frac{\Gamma[\frac{2+3(n-1)}{4}]}{\Gamma[\frac{1+3n}{4}]}\left(\frac{z_t}{z_h}\right)^{3n}\right]~.
\end{eqnarray}
This in the IR limit yields  
\begin{eqnarray}
T_{ent}= \frac{2}{3} T_H.
\end{eqnarray}
This clearly shows that the definition in \cite{jprl} for the entanglement temperature yields the Hawking temperature only upto a multiplicative constant factor in the IR limit. We shall further substantiate our claim by discussing the BTZ black hole case in the appendix of this paper.\\
We shall see in the next subsection that the definition of the generalized entanglement temperature $T_g$ given in this paper (eq.(\ref{betagen})) gives the exact Hawking temperature in the IR limit and also makes the entanglement temperature in the UV limit satisfy the thermodynamic like relation 
\begin{eqnarray}
E = \left(\frac{d-1}{d}\right)T_{e} ~S_{REE}^{(UV)}
\end{eqnarray}
where $T_{e}=T_g$ in the UV limit. This implies that the generalized thermodynamics like law 
\begin{eqnarray}
E = \left(\frac{d-1}{d}\right)T_{g} S_{REE}
\end{eqnarray}
becomes an exact relation in the entire domain of the subsystem of length $l$.\\
\noindent The analogue of the above relation for the BTZ black hole case reads \cite{cpark3} 
\begin{eqnarray}
E= \frac{1}{2}T_g~S_{REE}~.
\end{eqnarray}
Here also we find that there is a factor of $1/2$ on the right hand side of the above equation. This factor ensures $T_g$ to reproduce the exact Hawking temperature of the BTZ black hole in the IR ($l\rightarrow\infty$) limit.
      
\subsection{Behaviour of $\beta_g$ in the IR region}
In this subsection, we shall study the behaviour of the generalized entanglement temprerature $T_g$ in the IR region of the theory. Before going to the IR regime, we shall simplify the expression of $S_{REE}$ given in eq.(\ref{renorm}). This gives
\begin{eqnarray}\label{reemod}
S_{REE} &=& \frac{2^{(d-3)}\pi^{\frac{d-1}{2}}}{(d-2)G_{d+1}}\bigg(\frac{\Gamma[\frac{d}{2(d-1)}]}{\Gamma[\frac{1}{2(d-1)}]}\bigg)^{(d-1)} \bigg(\frac{L}{l}\bigg)^{(d-2)} - \frac{\sqrt{\pi}}{2G_{d+1}} \bigg(\frac{L}{z_t}\bigg)^{(d-2)} \frac{\Gamma[\frac{d}{2(d-1)}]}{(d-2)\Gamma[\frac{1}{2(d-1)}]}\nonumber\\
&&+\frac{1}{2G_{d+1}} \bigg(\frac{L}{z_t}\bigg)^{(d-2)} \sum_{n=1}^{\infty} \bigg[1+\frac{d-1}{2+(n-1)d}\bigg]\bigg(\frac{1}{1+nd}\bigg) \frac{\Gamma[n+\frac{1}{2}]}{\Gamma[n+1]}\frac{\Gamma[\frac{d(1+n)}{2(d-1)}]}{\Gamma[\frac{1+nd}{2(d-1)}]} \bigg(\frac{z_t}{z_h}\bigg)^{nd}\nonumber\\
\end{eqnarray}
where we have used the identity $\Gamma[p+1]=p\Gamma[p]$. By using eq.(\ref{system}) we obtain
\begin{eqnarray}
\frac{l}{2z_t}-\sqrt{\pi}\frac{\Gamma[\frac{d}{2(d-1)}]}{\Gamma[\frac{1}{2(d-1)}]} = \sum_{n=1}^{\infty} \bigg(\frac{1}{1+nd}\bigg) \frac{\Gamma[n+\frac{1}{2}]}{\Gamma[n+1]}\frac{\Gamma[\frac{d(1+n)}{2(d-1)}]}{\Gamma[\frac{1+nd}{2(d-1)}]} \bigg(\frac{z_t}{z_h}\bigg)^{nd}~.
\end{eqnarray}
Substituting this in eq.(\ref{reemod}) gives
\begin{eqnarray}
S_{REE}&=& \frac{1}{4G_{d+1}}\frac{L^{d-2}l}{z_t^{d-1}}+\frac{1}{2G_{d+1}}\bigg(\frac{L}{z_t}\bigg)^{(d-2)} \sum_{n=1}^{\infty} \bigg[\frac{d-1}{2+(n-1)d}\bigg]\bigg(\frac{1}{1+nd}\bigg) \frac{\Gamma[n+\frac{1}{2}]}{\Gamma[n+1]}\frac{\Gamma[\frac{d(1+n)}{2(d-1)}]}{\Gamma[\frac{1+nd}{2(d-1)}]} \bigg(\frac{z_t}{z_h}\bigg)^{nd}\nonumber\\
&&+\frac{2^{(d-3)}\pi^{\frac{d-1}{2}}}{(d-2)G_{d+1}}\bigg(\frac{\Gamma[\frac{d}{2(d-1)}]}{\Gamma[\frac{1}{2(d-1)}]}\bigg)^{(d-1)} \bigg(\frac{L}{l}\bigg)^{(d-2)}- \frac{\sqrt{\pi}}{2G_{d+1}} \bigg(\frac{L}{z_t}\bigg)^{(d-2)}\frac{(d-1)}{(d-2)} \frac{\Gamma[\frac{d}{2(d-1)}]}{\Gamma[\frac{1}{2(d-1)}]}~.
\end{eqnarray}
Now we shall take the IR limit in the above expression by taking the limit $z_t \rightarrow z_h$. In this limit, the infinite series in the above equation goes as $\approx \frac{1}{n^2}(\frac{z_t}{z_h})^{nd}$ for large $n$, which implies that the series converges in the $z_t\rightarrow z_h$ limit. The expression for $S_{REE}$ in the IR limit reads
\begin{eqnarray}\label{irrenorm}
S_{REE}&=& \frac{1}{4G_{d+1}}\frac{L^{d-2}l}{z_h^{d-1}}+\frac{1}{2G_{d+1}}\bigg(\frac{L}{z_h}\bigg)^{(d-2)} \sum_{n=1}^{\infty} \bigg[\frac{d-1}{2+(n-1)d}\bigg]\bigg(\frac{1}{1+nd}\bigg) \frac{\Gamma[n+\frac{1}{2}]}{\Gamma[n+1]}\frac{\Gamma[\frac{d(1+n)}{2(d-1)}]}{\Gamma[\frac{1+nd}{2(d-1)}]} \nonumber\\
&+&\frac{2^{(d-3)}\pi^{\frac{d-1}{2}}}{(d-2)G_{d+1}}\bigg(\frac{\Gamma[\frac{d}{2(d-1)}]}{\Gamma[\frac{1}{2(d-1)}]}\bigg)^{(d-1)} \bigg(\frac{L}{l}\bigg)^{(d-2)}- \frac{\sqrt{\pi}}{2G_{d+1}} \bigg(\frac{L}{z_h}\bigg)^{(d-2)} \frac{(d-1)}{(d-2)}\frac{\Gamma[\frac{d}{2(d-1)}]}{\Gamma[\frac{1}{2(d-1)}]}~.
\end{eqnarray}
The above result can be recast in the form 
\begin{eqnarray}\label{corrected}
S_{REE} &=& \frac{\bar{A}}{4G_{d+1}}\bigg[1 +  \Delta_1 \bigg(\frac{z_h}{l}\bigg) + \Delta_2 \bigg(\frac{z_h}{l}\bigg)^{d-1}\bigg]\nonumber \\
&=& S_{th}\bigg[1 +  \Delta_1 \bigg(\frac{z_h}{l}\bigg) + \Delta_2 \bigg(\frac{z_h}{l}\bigg)^{d-1}\bigg]
\end{eqnarray}
where $\bar{A} \approx \frac{L^{d-2}l}{z_h^{d-1}}$ is the area of the of the near horizon part of the static minimal surface in the IR limit and $\Delta_1$ and $\Delta_2$ is given by 
\begin{eqnarray}
\Delta_1 &=& 2 \sum_{n=1}^{\infty} \bigg[\frac{d-1}{2+(n-1)d}\bigg]\bigg(\frac{1}{1+nd}\bigg) \frac{\Gamma[n+\frac{1}{2}]}{\Gamma[n+1]}\frac{\Gamma[\frac{d(1+n)}{2(d-1)}]}{\Gamma[\frac{1+nd}{2(d-1)}]}- 2\sqrt{\pi}  \frac{(d-1)}{(d-2)}\frac{\Gamma[\frac{d}{2(d-1)}]}{\Gamma[\frac{1}{2(d-1)}]}\\
\Delta_2 &=& \frac{2^{(d-1)}\pi^{\frac{d-1}{2}}}{(d-2)}\bigg(\frac{\Gamma[\frac{d}{2(d-1)}]}{\Gamma[\frac{1}{2(d-1)}]}\bigg)^{(d-1)}.
\end{eqnarray}
\noindent Substituting eq.(\ref{corrected}) in eq.(\ref{betagen}), we obtain the behaviour of $T_g$ in the IR region to be
	
	\begin{eqnarray}\label{betair}
	\beta_g \equiv \frac{1}{T_{g}} &=&  \frac{4\pi z_h}{d}\bigg[1 +  \Delta_1 \bigg(\frac{z_h}{l}\bigg) + \Delta_2 \bigg(\frac{z_h}{l}\bigg)^{d-1} \bigg]\nonumber \\
	&=& \frac{1}{T_H}\bigg[1+ \mathrm{Correction~terms} \bigg]~. 
	\end{eqnarray}
	It is interesting to notice that in the IR regime the leading term of the generalized entanglement temperature $\beta_g^{-1}$ is equal to the inverse of the Hawking temperature. The rest of the terms are correction terms and subsystem size dependent. In the large $l$ limit,  these correction terms are smaller in amplitude compared to the Hawking temperature $T_H$ which is $l$ independent and the correction terms are inversely proportional to $l$.\\
For $d=3$ corresponding to the $\mathrm{AdS}_{3+1}$ black hole eq(s).(\ref{corrected}), \ref{betair} read
\begin{eqnarray}
S_{REE}&=& S_{th}\bigg[1 -1.7605 \bigg(\frac{z_h}{l}\bigg) + 1.4355 \bigg(\frac{z_h}{l}\bigg)^{2}\bigg]
\end{eqnarray}
\begin{eqnarray}
\beta_g \equiv \frac{1}{T_{g}} &=& \frac{1}{T_{H}}\bigg[1 -1.7605 \bigg(\frac{z_h}{l}\bigg) + 1.4355 \bigg(\frac{z_h}{l}\bigg)^{2}\bigg]
\end{eqnarray}
where
\begin{eqnarray}
S_{th} = \frac{1}{4G}\frac{Ll}{z_h^{2}}~;~T_H=\frac{3}{4\pi z_h}~.
\end{eqnarray}
Similarly, for $d=4$ corresponding to $\mathrm{AdS}_{4+1}$ black hole eq(s).(\ref{corrected}), \ref{betair} read

\begin{eqnarray}
S_{REE}&=& S_{th} \bigg[1 -0.6658 \bigg(\frac{z_h}{l}\bigg) + 0.3206 \bigg(\frac{z_h}{l}\bigg)^{3}\bigg]
\end{eqnarray}
\begin{eqnarray}
\beta_g \equiv \frac{1}{T_{g}} &=& \frac{1}{T_{H}}\bigg[1 -0.6658 \bigg(\frac{z_h}{l}\bigg) + 0.3206 \bigg(\frac{z_h}{l}\bigg)^{3}\bigg]
\end{eqnarray}
where
\begin{eqnarray}
S_{th} = \frac{1}{4G}\frac{L^2l}{z_h^{3}}~;~T_H=\frac{1}{\pi z_h}~.
\end{eqnarray}	
Note that although in the IR limit, we use the fact $z_t \approx z_h$, however it is known that in the high temperature limit (IR limit), the static minimal surface approaches the event horizon but always stays at a finite distance behind the event horizon $z_h$ \cite{hubeny2}.
The above expressions show that in the IR region the leading term of $S_{REE}$ is the amount of Bekenstein-hawking entropy contained in the boundary subsystem volume given in eq.(\ref{hawking}).\\

\subsection{Behaviour of $\beta_g$ in the UV region}
In this subsection, we look at the UV behaviour of the generalized entanglement temperature $T_g$. In the UV region ($\frac{z_t}{z_h}\ll1, \frac{l}{z_h}\ll1$) the renormalized holographic entanglement entropy given in eq.(\ref{renorm}) reads
\begin{eqnarray}\label{UVrenorm}
S_{REE}&\approx&\frac{2^{(d-3)}\pi^{\frac{d-1}{2}}}{(d-2)G_{d+1}}\bigg(\frac{\Gamma[\frac{d}{2(d-1)}]}{\Gamma[\frac{1}{2(d-1)}]}\bigg)^{(d-1)} \bigg(\frac{L}{l}\bigg)^{(d-2)} -~ \frac{\sqrt{\pi}}{2(d-2)G_{d+1}} \frac{\Gamma[\frac{d}{2(d-1)}]}{\Gamma[\frac{1}{2(d-1)}]} \bigg(\frac{L}{z_t}\bigg)^{(d-2)}\nonumber\\ 
&&+\frac{\sqrt{\pi}}{8(d-1)G_{d+1}} \frac{L^{d-2}}{z_h^d} \frac{\Gamma[\frac{1}{d-1}]}{\Gamma[\frac{d+1}{2(d-1)}]}~ z_t^2+\frac{\sqrt{\pi}}{16G_{d+1}} \frac{L^{d-2}}{z_h^{2d}} \frac{\Gamma[\frac{2+d}{2(d-1)}]}{\Gamma[\frac{2}{2(d-1)}]}~ z_t^{d+2}. 
\end{eqnarray}
The subsystem size in the UV limit reads
\begin{eqnarray}\label{luv}
l\approx2z_t \sqrt{\pi} \frac{\Gamma[\frac{d}{2(d-1)}]}{\Gamma[\frac{1}{2(d-1)}]} +\frac{z_t\sqrt{\pi}}{(d^2-1)} \frac{\Gamma[\frac{1}{d-1}]}{\Gamma[\frac{d+1}{2(d-1)}]} \bigg(\frac{z_t}{z_h}\bigg)^d+\frac{z_t\sqrt{\pi}}{4}\frac{(d+2)}{(1+2d)}\frac{\Gamma[\frac{d+2}{2(d-1)}]}{\Gamma[\frac{3}{2(d-1)}]}\left(\frac{z_t}{z_h}\right)^{2d}~. 
\end{eqnarray}
Inverting eq.(\ref{luv}), we obtain
\begin{eqnarray}\label{ztuv}
z_t &\approx& \frac{l}{2\sqrt{\pi}}  \frac{\Gamma[\frac{1}{2(d-1)}]}{\Gamma[\frac{d}{2(d-1)}]} \bigg[1-\frac{1}{2^{d+1}(d^2-1)\pi^{d/2}}\frac{\Gamma[\frac{1}{d-1}]}{\Gamma[\frac{d+1}{2(d-1)}]}\left(\frac{l}{z_h}\right)^{d} \bigg(\frac{\Gamma[\frac{1}{2(d-1)}]}{\Gamma[\frac{d}{2(d-1)}]}\bigg)^{d+1}\nonumber\\
&&-\frac{1}{2^{2d+3}\pi^{d}}\frac{(d+2)}{(1+2d)}\frac{\Gamma[\frac{d+2}{2(d-1)}]}{\Gamma[\frac{3}{2(d-1)}]}\left(\frac{l}{z_h}\right)^{2d}\bigg(\frac{\Gamma[\frac{1}{2(d-1)}]}{\Gamma[\frac{d}{2(d-1)}]}\bigg)^{2d+1}\bigg].
\end{eqnarray}
Now substituting eq.(\ref{ztuv}) in eq.(\ref{UVrenorm}), we get

\begin{eqnarray}\label{uvs}
	S_{REE} &=& \frac{L^{d-2}l^2}{32 \sqrt{\pi} G_{d+1} z_h^d (d+1)} \frac{\Gamma[\frac{1}{d-1}]}{\Gamma[\frac{d+1}{2(d-1)}]} \bigg[\frac{\Gamma[\frac{1}{2(d-1)}]}{\Gamma[\frac{d}{2(d-1)}]}\bigg]^2\times\left[1+\xi\left(\frac{l}{z_h}\right)^d+...\right]\nonumber\\
	&\equiv& S_{REE}^{(UV)} \times\left[1+\xi\left(\frac{l}{z_h}\right)^d+...\right]
\end{eqnarray}
where
\begin{eqnarray}
\xi=\frac{(d^2-1)}{2^{d+1}(1+2d)\pi^{d/2}}\left(\frac{\Gamma[\frac{1}{2(d-1)}]}{\Gamma[\frac{d}{2(d-1)}]}\right)^d\left(\frac{\Gamma[\frac{d+2}{2(d-1)}]}{\Gamma[\frac{3}{2(d-1)}]}-\frac{2(1+2d)}{(d^2-1)}\left(\frac{\Gamma[\frac{d}{d-1}]}{\Gamma[\frac{d+1}{2(d-1)}]}\right)^2\frac{\Gamma[\frac{1}{2(d-1)}]}{\Gamma[\frac{d}{2(d-1)}]}\right)~.
\end{eqnarray}
The leading term $S_{REE}^{(UV)}$ in eq.(\ref{uvs}) matches with the result obtained in \cite{jprl}. The next term captures the subleading effects in the UV domain of the dual field theory.\\
\noindent The generalized entanglement temperature $T_g$ in the UV limit reads
\begin{eqnarray}\label{uvt}
\beta_g \equiv \frac{1}{T_g}&=& \frac{\sqrt{\pi}l}{2d(d+1)}\frac{\Gamma[\frac{1}{d-1}]}{\Gamma[\frac{d+1}{2(d-1)}]} \bigg[\frac{\Gamma[\frac{1}{2(d-1)}]}{\Gamma[\frac{d}{2(d-1)}]}\bigg]^2\times\left[1+\xi\left(\frac{l}{z_h}\right)^d+...\right]\nonumber\\
&\equiv&\frac{1}{T_{e}}\left[1+\xi\left(\frac{l}{z_h}\right)^d+...\right]
\end{eqnarray}
where
\begin{eqnarray}\label{tentnew}
T_{e}\equiv T_{g}^{(UV)}= \frac{2d(d+1)}{\sqrt{\pi}l}\frac{\Gamma[\frac{d+1}{2(d-1)}]}{\Gamma[\frac{1}{d-1}]}\left[\frac{\Gamma[\frac{d}{2(d-1)}]}{\Gamma[\frac{1}{2(d-1)}]}\right]^2~.
\end{eqnarray}
 The leading term corresponds to the generalized entanglement temperature in the UV limit and the second term captures the subleading effects in the UV domain. It is interesting to observe that in the UV limit, the leading term of the generalized entanglement temperature $T_g$ have the same characteristics as the entanglement temperature $T_{ent}$ obtained in \cite{jprl}, as both are inversely proportional to the subsystem size $l$ as expected from dimensional grounds. At this point, we would also like to mention that $T_{ent}$ defined in \cite{jprl} reduces to the following expression in the UV limit \cite{jprl}
 \begin{eqnarray}\label{tentold}
 T_{ent}= \frac{2(d^2-1)}{\sqrt{\pi}l}\frac{\Gamma[\frac{d+1}{2(d-1)}]}{\Gamma[\frac{1}{d-1}]}\left[\frac{\Gamma[\frac{d}{2(d-1)}]}{\Gamma[\frac{1}{2(d-1)}]}\right]^2
 \end{eqnarray}
 where only the leading term has been kept. \\
 From eq.(s)(\ref{tentnew}),(\ref{tentold}), we observe that $T_e$ is related to $T_{ent}$ as
 \begin{eqnarray}
 T_e = \left(\frac{d}{d-1}\right)~T_{ent}~.
 \end{eqnarray} 
  Finally, it is reassuring to note that the thermodynamics like relation 
 \begin{eqnarray}
 E= \left(\frac{d-1}{d}\right)T_{e} ~S_{REE}^{(UV)}
 \end{eqnarray}
 is satisfied.\\
For $d=3$ corresponding to the AdS$_{3+1}$ black hole,  eq.(s).(\ref{uvs}),(\ref{uvt}) read
\begin{eqnarray}
S_{REE}&=&S_{REE}^{(UV)} \times\left[1-\frac{1}{14\pi^{3/2}}\left(\frac{\Gamma[\frac{1}{4}]}{\Gamma[\frac{3}{4}]}\right)^3\left(\frac{7\pi\Gamma[\frac{1}{4}]}{16\Gamma[\frac{3}{4}]}-\frac{\Gamma[\frac{5}{4}]}{\Gamma[\frac{3}{4}]}\right) \left(\frac{l}{z_h}\right)^3+...\right]\nonumber\\
&=& S_{REE}^{(UV)} \times\left[1-1.10529 \left(\frac{l}{z_h}\right)^3+...\right]
\end{eqnarray}
\begin{eqnarray}
\frac{1}{T_g}&=& \frac{1}{T_{e}}\times\left[1-\frac{1}{14\pi^{3/2}}\left(\frac{\Gamma[\frac{1}{4}]}{\Gamma[\frac{3}{4}]}\right)^3\left(\frac{7\pi\Gamma[\frac{1}{4}]}{16\Gamma[\frac{3}{4}]}-\frac{\Gamma[\frac{5}{4}]}{\Gamma[\frac{3}{4}]}\right) \left(\frac{l}{z_h}\right)^3+...\right]\nonumber\\
&=&\frac{1}{T_{e}}\times\left[1-1.10529 \left(\frac{l}{z_h}\right)^3+...\right]
\end{eqnarray}
where
\begin{eqnarray}
S_{REE}^{(UV)}= \frac{L l^2}{128G_4 z_h^3}\left(\frac{\Gamma[\frac{1}{4}]}{\Gamma[\frac{3}{4}]}\right)^2~;~T_{e} =  \frac{24}{\pi l}\left(\frac{\Gamma[\frac{3}{4}]}{\Gamma[\frac{1}{4}]}\right)^2~.
\end{eqnarray}
Similarly, for $d=4$ corresponding to $\mathrm{AdS}_{4+1}$ black hole, eq(s).(\ref{uvs}), \ref{uvt} read
\begin{eqnarray}
S_{REE}&=&S_{REE}^{(UV)} \times\left[1-\frac{5}{96\pi^{2}}\left(\frac{\Gamma[\frac{1}{6}]}{\Gamma[\frac{2}{3}]}\right)^4\left(\frac{6\Gamma[\frac{1}{6}]\Gamma[\frac{4}{3}]^2}{5\Gamma[\frac{2}{3}]\Gamma[\frac{5}{6}]^2}-\frac{1}{\sqrt{\pi}}\right) \left(\frac{l}{z_h}\right)^4+...\right]\nonumber\\
&=& S_{REE}^{(UV)} \times\left[1-3.80143\left(\frac{l}{z_h}\right)^4+...\right]
\end{eqnarray}
\begin{eqnarray}
\frac{1}{T_g}&=& \frac{1}{T_{e}}\times\left[1-\frac{5}{96\pi^{2}}\left(\frac{\Gamma[\frac{1}{6}]}{\Gamma[\frac{2}{3}]}\right)^4\left(\frac{6\Gamma[\frac{1}{6}]\Gamma[\frac{4}{3}]^2}{5\Gamma[\frac{2}{3}]\Gamma[\frac{5}{6}]^2}-\frac{1}{\sqrt{\pi}}\right) \left(\frac{l}{z_h}\right)^4+...\right]\nonumber\\
&=& \frac{1}{T_{e}}\times\left[1-3.80143\left(\frac{l}{z_h}\right)^4+...\right]
\end{eqnarray}
where
\begin{eqnarray}
S_{REE}^{(UV)}= \frac{\Gamma[\frac{1}{6}]^2\Gamma[\frac{1}{3}]L^2l^2}{160\sqrt{\pi}G_5 z_h^4\Gamma[\frac{2}{3}]^2\Gamma[\frac{5}{6}]}~;~T_{e} =  \frac{40}{\sqrt{\pi} l}\frac{\Gamma[\frac{2}{3}]^2\Gamma[\frac{5}{6}]}{\Gamma[\frac{1}{6}]^2\Gamma[\frac{1}{3}]}~.
\end{eqnarray}

\section{Behaviour of $\beta_g$ with respect to the subsystem size of length $l$}\label{sec3}
\begin{wrapfigure}[17]{l}{0.55\textwidth}
	\centering
	\includegraphics[width=0.55\textwidth]{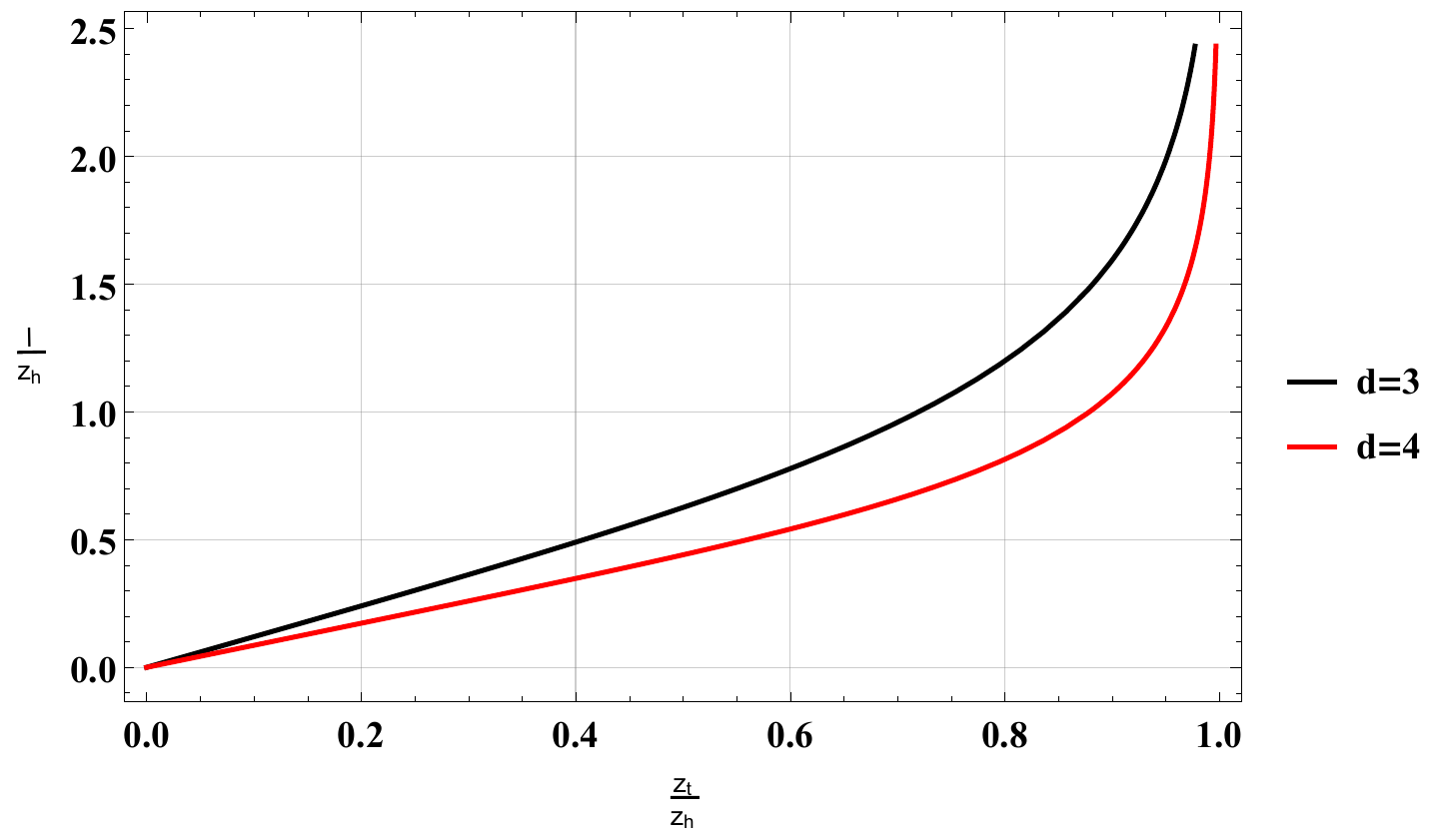}
	\caption{\textbf{Variation of the subsystem size $\frac{l}{z_h}$ with respect to $\frac{z_t}{z_h}$}.}
	\label{ztvsl}
\end{wrapfigure}
In this section, we set $d=3,4$ to observe the behavior of $\frac{\beta_g}{z_h}$ with respect to $\frac{l}{z_h}$ in $AdS_{3+1}$ and $AdS_{4+1}$ respectively. The strategy we take to do the study is quite simple. It can be observed that the subsystem size $l$ given in eq.(\ref{system}) is function of the turning point $z_t$ alone (treating $z_h$ as a constant). Substituting $l(z_t)$ in eq.(\ref{renorm}), we can obtain $S_{REE}(z_t)$.
This means that both $l$ and the renormalized holographic entanglement entropy $S_{REE}$ are functions of the turning point $z_t$ alone. Hence, the generalized entanglement temperature $T_g$ is also a function of the turning point $z_t$ only, symbolically $\beta_g(z_t)$.   
In Fig.(\ref{ztvsl}), we make use of eq.(\ref{system}) to show the variation of the boundary subsystem size $l$ with respect to the turning point in the bulk $z_t$ for $AdS_{3+1}$ and $AdS_{4+1}$. However, to avoid dimensionful quantities, we have plotted the dimensionless quantities, namely, $\frac{l}{z_h}$ against $\frac{z_t}{z_h}$. It can be seen that the nature of the curve remains the same in higher dimensions and in both cases $\frac{l}{z_h}$ diverges in the limit $\frac{z_t}{z_h}\rightarrow1$. We then compute $\frac{d\left(\frac{\beta_g}{z_h}\right)}{d\left(\log \left(\frac{l}{z_h}\right)\right)}$ for any dimension $d\geq 3$ and plot it against the dimensionless quantity $\frac{l}{z_h}$. This has been done to show the crossover from UV region to the IR region in the dual field theory. Interestingly, we observe a critical length $l_c$ in units of $z_h$ from the plot which depends on the dimensions of the spacetime geometry. It is also interesting to note that the variation of the subsystem size $\frac{l}{z_h}$ with $\frac{z_t}{z_h}$ has a similar nature in $d=2$ case corresponding to the BTZ black hole. This is quite interesting because the $d=2$ case is very much different from the higher dimensional ($d \geq 3$) black holes.\\
 \subsection{$AdS_{3+1}$ black hole}
\begin{figure}[h!]
	\begin{minipage}[t]{0.95\textwidth}
		\centering\includegraphics[width=0.8\textwidth]{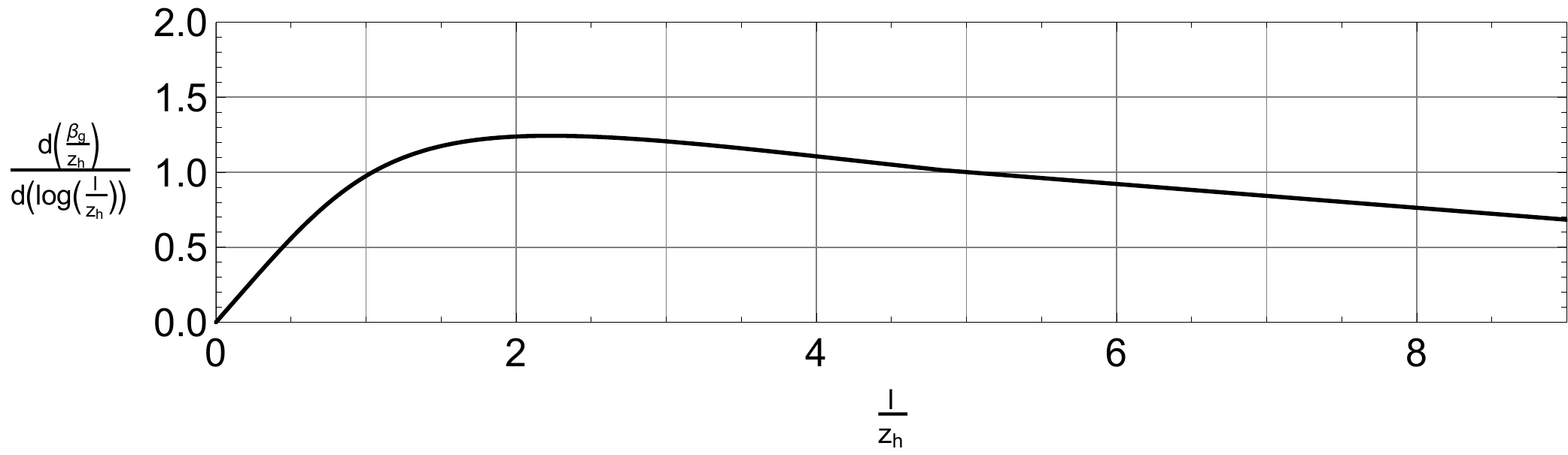}\\
	\end{minipage}
	\caption{\textbf{Variation of $\frac{d\left(\frac{\beta_g}{z_h}\right)}{d\left(\log \left(\frac{l}{z_h}\right)\right)}$ with respect to $\frac{l}{z_h}$  in $AdS_{3+1}$ black hole}.}
	\label{betaflowd3}
\end{figure}

  \
   \begin{wrapfigure}[13]{r}{0.6\textwidth}
  	\centering\includegraphics[width=0.6\textwidth]{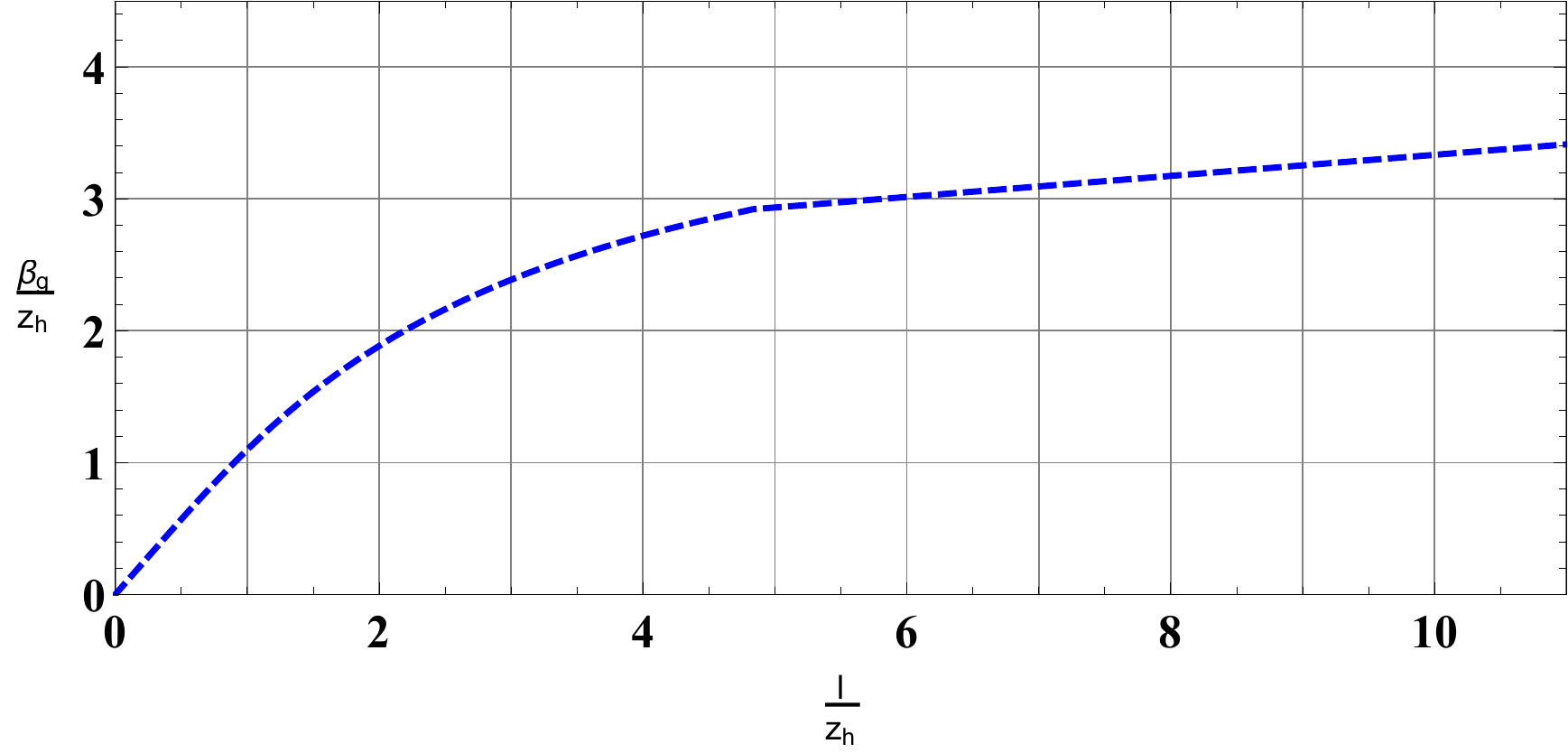}\\
  	\caption{\textbf{Variation of $\frac{\beta_g}{z_h}$ with respect to $\frac{l}{z_h}$ in $AdS_{3+1}$ black hole.}}
  	\label{betad3}
\end{wrapfigure}
 \noindent Substituting $d=3$ in eq.(s)(\ref{renorm}) and (\ref{system}) we obtain the renormalized entanglement entropy $S_{REE}(z_t)$ and $l(z_t)$ for $AdS_{3+1}$ black hole. Further, substituting $d=3$ in eq.(\ref{betagen}) gives us $\beta_{g}(z_t)$. Fig.(\ref{betad3}) represents the variation of $\frac{\beta_g}{z_h}$ with respect $\frac{l}{z_h}$ in AdS$_{3+1}$ spacetime. From the plot we observe that $\beta_g$ approaches the inverse of Hawking temperature in the large $l$ limit. In Fig.(\ref{betaflowd3}) we represent the variation of $\frac{d\left(\frac{\beta_g}{z_h}\right)}{d\left(\log \left(\frac{l}{z_h}\right)\right)}$ with respect to the dimensionless quantity $\frac{l}{z_h}$ for the $AdS_{3+1}$ black hole. The flow shows a maximum at $l_c \approx 2.031 z_h$ and after that it asymptotically decays which represents the thermalization of the excitations occuring in the boundary field theory. This maximum value point can be interpreted as a crossover scale. Since upto $l_c$, the quantum mechanical behavior dominates and after that the quantum mechanical excitations in the boundary field theory starts to thermalize as we move towards the IR limit which makes the flow to decay asymptotically and reveals the thermal nature of the system.
\\
\subsection{\textbf{$AdS_{4+1}$} black hole: $\mathcal{N}=4$ Super Yang-Mills theory}
\begin{figure}[!h]
 	\begin{minipage}[t]{1\textwidth}
 		\centering\includegraphics[width=0.8\textwidth]{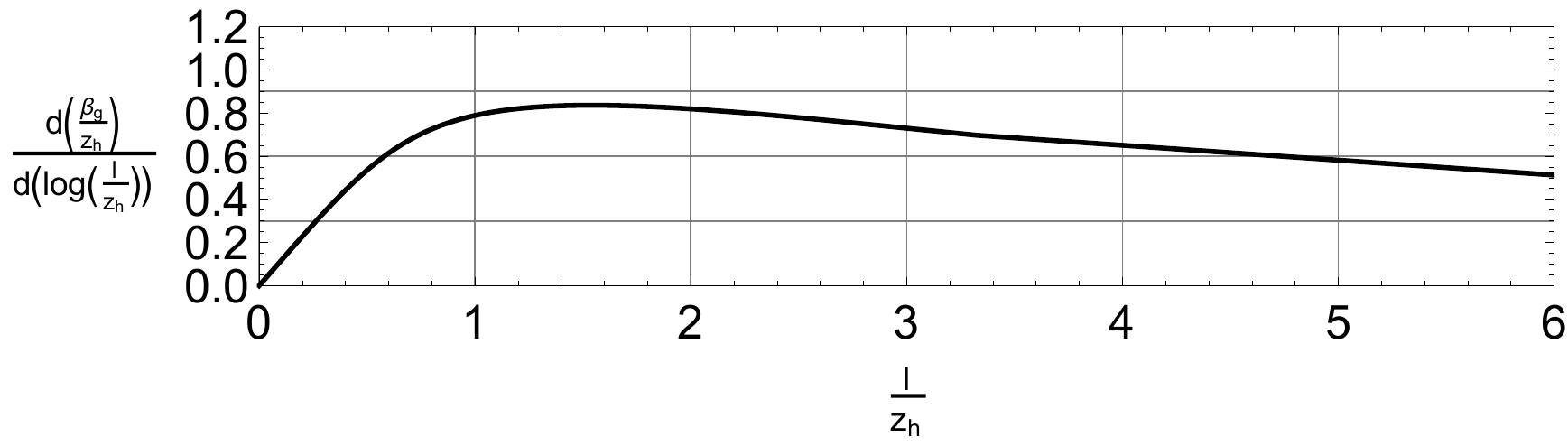}\\
 	\end{minipage}
 	\caption{\textbf{Variation of $\frac{d\left(\frac{\beta_g}{z_h}\right)}{d\left(\log \left(\frac{l}{z_h}\right)\right)}$ with respect to $\frac{l}{z_h}$ in $AdS_{4+1}$ black hole}.}
 	\label{betflowad4}
 \end{figure}

This scenario is quite interesting as in this case the ground state HEE corresponds to the EE of $\mathcal{N}=4$ Super yang-mills in $(3+1)$-dimensions which is dual to the pure $AdS_5$ spacetime. 
 \begin{wrapfigure}[13]{r}{0.6\textwidth}
 	\centering\includegraphics[width=0.6\textwidth]{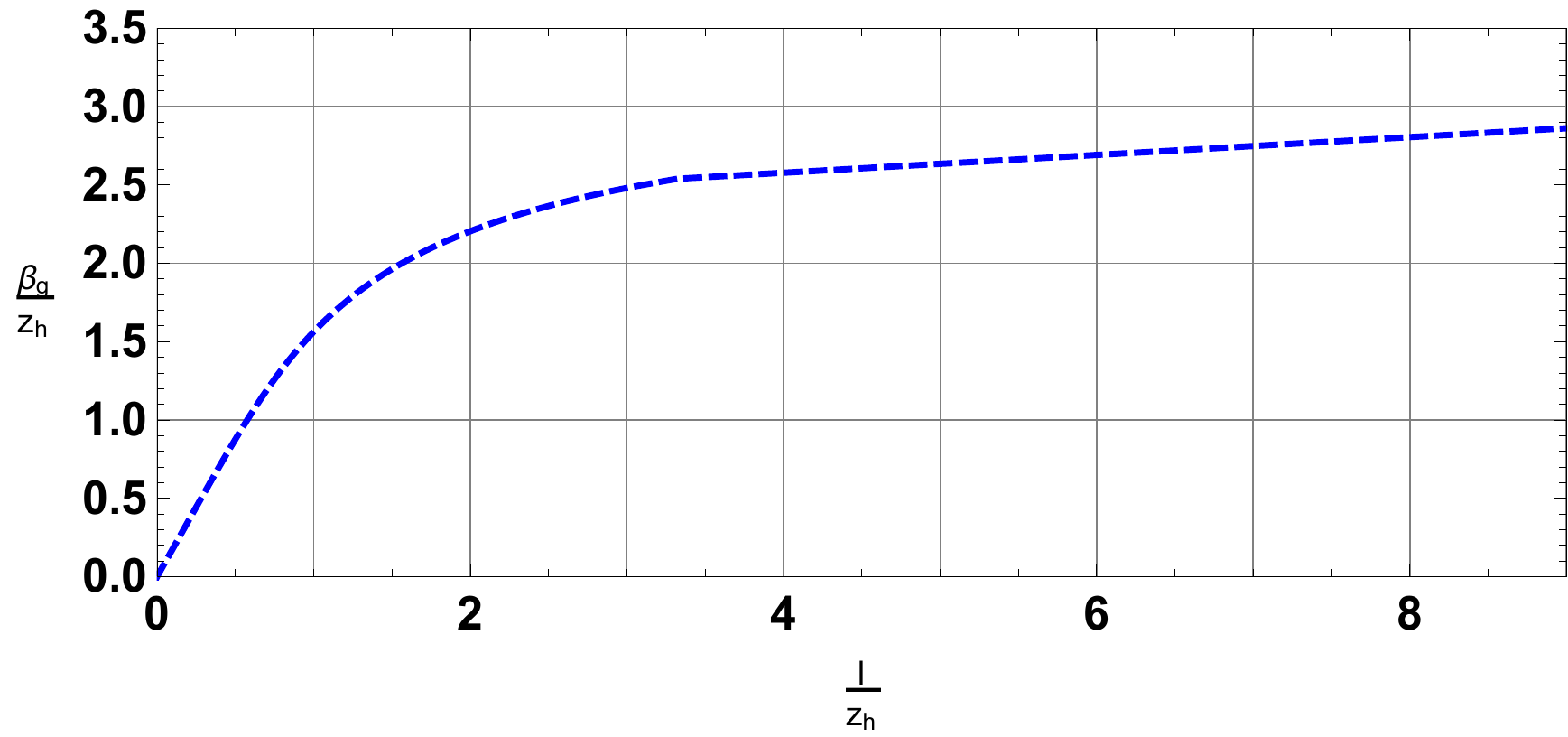}
 	\caption{\textbf{Variation of $\frac{\beta_g}{z_h}$ with respect to $\frac{l}{z_h}$ in $AdS_{4+1}$ black hole.}}
 	\label{betad4}
 \end{wrapfigure}   
 \noindent  The emergence of the $AdS_5$ bulk dual from the EE of $\mathcal{N}=4$ SYM has been studied in \cite{bulk}. Once again we graphically represent the variation of $\frac{\beta_g}{z_h}$ with $\frac{l}{z_h}$ in Fig.(\ref{betad4}) and $\frac{d\left(\frac{\beta_g}{z_h}\right)}{d\left(\log \left(\frac{l}{z_h}\right)\right)}$ in Fig.(\ref{betflowad4}).
  From Fig.(\ref{betad4}) we again observe that $\beta_g$ approaches inverse of the Hawking temperature in the large $l$ limit. In Fig.(\ref{betflowad4}) the quantum mechanical to thermal crossover is at $l_c \approx 1.45z_h$ where the flow shows the maximum value, after that it moves towards the IR regime and the quantum mechanical excitations in the boundary $\mathcal{N}=4$ SYM theory starts to thermalize. This plot shows that in higher dimension the thermalization in the boundary field theory occurs at smaller value of the subsytem size $l$. The plots once again reveal that the generalized entanglement temperature $T_g$ firmly represents the behavior of the renormalized holographic entanglement along the whole scale (subsystem size) of the theory.

 \section{Conclusion}\label{sec4}   
We summarize our findings now. In this paper, we holographically compute the entanglement entropy of both the ground state and the excited state of a $d$-dimensional boundary conformal field theory living in the boundary of a $(d+1)$-dimensional $AdS$ space. By utilizing these results of entanglement entropy we define a divergence free renormalized entanglement entropy $(S_{REE})$ in dimensions $d \geq 3$. With the help of the first law of black hole thermodynamics and $S_{REE}$, we then define a generalized entanglement temperature $T_g$, defined in such a way so as to reproduce the Hawking temperature of the black hole in the infrared limit. This definition of the generalized entanglement temperature is valid in both the ultraviolet and infrared domain of the theory unlike the entanglement temperature $T_{ent}$ (defined in \cite{jprl}) of entanglement thermodynamics computed by Fefferman-Graham expansion of the concerned metric \cite{alishahiha} in the ultraviolet domain. Our definition of the generalized entanglement temperature is different from that given in \cite{jprl}. The definition given in \cite{jprl} matches with the Hawking temperature of the black hole only upto a constant multiplicative factor in the infrared limit. We have shown this explicitly for the SAdS$_{4}$ black hole and the BTZ black hole. Further, our definition of the generalized entanglement temperature satisfies the thermodynamics like law $E=\left(\frac{d-1}{d}\right) T_{g}~S_{REE}$  and holds in the entire length scale $(l)$ of the dual field theory. This generalized entanglement temperature should not be identified as the thermodynamic temperature of the CFT which corresponds to the Hawking temperature of the black hole. We then study the behaviour of $T_g$ in the ultraviolet and infrared domain of the theory. In the IR regime, $T_g$ gives the inverse Hawking temperature along with some correction terms depending upon the subsystem size $l$ of the boundary field theory. In the ultraviolet regime, $T_g$ is inversely proportional to the subsystem size $l$ and gives the entanglement temperature $T_{e}$. We find that $T_{e}$ also satisfies the thermodynamics like law $E=\left(\frac{d-1}{d}\right) T_{e}~S_{REE}^{(UV)}$ in the UV domain. We also observe that $T_e$ is related to $T_{ent}$ in \cite{jprl} as $T_e=\frac{d}{d-1}T_{ent}$. In the infrared domain, the renormalized holographic entanglement entropy $S_{REE}$ of the excited state of the boundary field theory thermalizes and yields the thermal entropy of the black hole. Interestingly, we observe that the leading correction to the thermal entropy is not logarithmic in nature for $d \geq 3$ which is in contrast to the BTZ black hole ($d=2$) case. In the ultraviolet domain, we obtain subleading corrections to the renormalized entanglement entropy $S_{REE}$ and $T_g$. We then provide explicit expressions including the subleading corrections for the AdS$_{3+1}$ and AdS$_{4+1}$ black holes. We then compute $\frac{d\left(\frac{\beta_g}{z_h}\right)}{d\left(\log \left(\frac{l}{z_h}\right)\right)}$ to capture the flow of the generalized entanglement temperature with respect to the dimensionless quantity $\frac{l}{z_h}$. We also graphically represent the flow of the generalized entanglement temperature  $\beta_g^{-1}$ along the whole length scale of the theory for $AdS_{3+1}$ and $AdS_{4+1}$ black holes. The analysis for the $\mathrm{AdS}_{4+1}$ is important in its own right since it corresponds to the $\mathcal{N}=4$ SYM theory at a finite temperature living in the boundary of the $\mathrm{AdS}_{4+1}$ black hole spacetime. The  flow of $\beta_g$ firmly shows quantum mechanical to thermal crossover of the theory. This crossover takes place at a particular size $l_c$ of the subsystem. Interestingly, we observe that $l_c$ decreases with increase in the spacetime dimensions. We find that $l_c \approx 2.031 z_h$ for $d=3$ and $l_c\approx 1.45 z_h$ for $d=4$ which is less than the value obtained for the $d=2$ case, namely, $l_c \approx 7.019$ \cite{cpark3}. This clearly indicates that the critical length decreases with the increase in the spacetime dimensions which in turn implies that the thermalization at the boundary CFT takes place much earlier in the subsystem size $l$. Furthermore, we also observe that the purely microscopic entanglement entropy evolves to a macroscopic thermal entropy revealing the microscopic origin of the real thermodynamic law in the context of black hole thermodynamics. We believe this connection between non-classical, microscopic entanglement thermodynamics and real thermodynamics is quite fascinating as it directly helps us to understand the microscopic origins of a macroscopic entity. The scenario discussed in this paper in principle can be extended for hyperscale violating spacetimes and also for a boundary field theory with a chemical potential. We leave these as future works.      
\section*{Appendix}
In this appendix we shall demonstrate for the BTZ black hole case that the generalized entanglement temperature defined as the ratio of the change in the internal energy to the renormalized entanglement entropy with an appropriate multiplicative factor reproduces the exact Hawking temperature of the BTZ black hole in the IR limit.\\
In case of the BTZ black hole, the thermal entropy $(S_{BH})$, change in internal energy $(E)$ and the Hawking temperature $(T_{H})$ satisfies the relation
\begin{eqnarray}
E = \frac{1}{2} T_H S_{BH}
\end{eqnarray} 
where $T_H$, $S_{BH}$ and $E$ are given by
\begin{eqnarray}
T_H = \frac{1}{2\pi z_h}~;~S_{BH} = \frac{l}{4G z_h}~;~E= \frac{l}{16\pi Gz_h^2}~.
\end{eqnarray} 
Motivated by the above form of the thermodynamic relation, the generalized entanglement temperature was defined in \cite{cpark3} as
\begin{eqnarray}
\frac{1}{T_g} = \frac{1}{2}\frac{S_{REE}}{E}
\end{eqnarray}
where $S_{REE}$ is the renormalized holographic entanglement entropy for the BTZ black hole given by
\begin{eqnarray}
S_{REE}&=& S_{BTZ}-S_{AdS_3}\nonumber\\
&=& \frac{1}{2G}\log\left[\left(\frac{2z_h}{l}\right)\sinh\left(\frac{l}{2z_h}\right)\right]~.
\end{eqnarray}
According to the prescription given in \cite{jprl} to compute the entanglement temperature, we have 
\begin{eqnarray}\label{prl}
T_{ent}=\frac{E}{S_{REE}}~.
\end{eqnarray}
In the UV domain ($\frac{l}{z_h}\ll1$), $S_{REE}$ can be expanded as
\begin{eqnarray}
S_{REE} = \frac{1}{48G}\left(\frac{l}{z_h}\right)^2\left[1-\frac{l^2}{120z_h^2}+...\right]~.
\end{eqnarray}
Therefore, eq.(\ref{prl}) gives the entanglement temperature of the BTZ black hole in the UV limit to be \cite{jprl}
\begin{eqnarray}
T_{ent}=\frac{3}{\pi l}~.
\end{eqnarray}
Now in the IR domain ($\frac{l}{z_h}\gg1$), $S_{REE}$ can be expanded as
\begin{eqnarray}
S_{REE} = \frac{1}{4G}\frac{l}{z_h}-\frac{1}{2G}\log\left(\frac{l}{4z_h}\right)+...~.
\end{eqnarray}
Once again using eq.(\ref{prl}), we obtain the entanglement temperature of the BTZ black hole in the IR limit to be
\begin{eqnarray}
T_{ent}&=&\frac{1}{4\pi z_h}+...\nonumber\\
&=&\frac{1}{2}T_H+...~.
\end{eqnarray}
Hence the prescription given in \cite{jprl} to compute the entanglement temperature matches with the Hawking temperature of the BTZ black hole upto a factor of $1/2$. We have also made a similar observation for the SAdS$_4$ black hole in section \ref{gent}.
 
\noindent On the other hand, using the relation given in eq.(\ref{betagen}), we obtain 
\begin{eqnarray}
T_g= \frac{6}{\pi l} 
\end{eqnarray}
in the UV domain, and 
\begin{eqnarray}
T_g=T_H+...
\end{eqnarray}
in the IR domain of the dual field theory.\\
\noindent In this paper we have generalized the definition of the entanglement temperature as
\begin{eqnarray}
\frac{1}{T_{g}}=\frac{(d-1)}{d}\frac{S_{REE}}{E}
\end{eqnarray}  
where the factor of $\left(\frac{d-1}{d}\right)$ has been chosen to reproduce the exact Hawking temperature of the black hole in the IR limit of the dual field theory.
\section*{Acknowledgements}
A.S. would like to acknowledge the support by Council of Scientific and Industrial Research (CSIR, Govt. of India) for Junior Research Fellowship. S.G. acknowledges the support of the Visiting Associateship programme of IUCAA, Pune. J.P.S. would like to acknowledge the support by University of Kalyani for Personal Research Grant. The authors would like to acknowledge the anonymous referee for very useful comments.


\end{document}